\documentclass[pra,showpacs]{revtex4}
\usepackage{amssymb}
\usepackage{amsmath}
\usepackage{graphicx}
\usepackage{makeidx}
\usepackage{subfigure}
\usepackage[english]{babel}
\usepackage{epsfig}

\begin{document}

\title{The ground state of binary systems with a periodic modulation of the
linear coupling}

\author{Armand Niederberger$^{1}$, Boris A. Malomed$^{2}$, and Maciej
Lewenstein$^{1,3}$}
\affiliation{$^{1}$ICFO -- Institut de
Ci\`{e}ncies Fot\`{o}niques, E-08860 Castelldefels
(Barcelona), Spain\\
$^{2}$Department of Physical Electronics, School of Electrical
Engineering, Faculty of Engineering, Tel Aviv University, Tel Aviv
69978,
Israel\\
$^{3}$ICREA -- Instituci\'{o} Catalana de Recerca i Estudis
Avan\c{c}ats, E-08010, Barcelona, Spain}

\begin{abstract}
We consider a quasi-one-dimensional two-component systm, described
by a pair of Nonlinear Schr\"{o}dinger/Gross-Pitaevskii Equations
(NLSEs/GPEs), which are coupled by the linear mixing, with local
strength $\Omega $, and by the nonlinear incoherent interaction. We
assume the self-repulsive nonlinearity in both components, and
include effects of a harmonic trapping potential. The model may be
realized in terms of periodically modulated slab waveguides in
nonlinear optics, and in Bose-Einstein condensates too. Depending on
the strengths of the linear and nonlinear couplings between the
components, the ground states (GSs) in such binary systems may be
symmetric or asymmetric. In this work, we introduce a periodic
spatial modulation of the linear coupling, making $\Omega $ an odd,
or even function of the coordinate. The sign flips of $\Omega (x)$
strongly modify the structure of the GS in the binary system, as the
relative sign of its components tends to lock to the local sign of
$\Omega $. Using a systematic numerical analysis, and an analytical
approximation, we demonstrate that the GS of the trapped system
contains one or several kinks (dark solitons) in one component,
while the other component does not change its sign. Final results
are presented in the form of maps showing the number of kinks in the
GS as a function of the system's parameters, with the odd/even
modulation function giving rise to the odd/even number of the kinks.
The modulation of $\Omega (x)$ also produces a strong effect on the
transition between states with nearly equal and strongly unequal
amplitudes of the two components.
\end{abstract}


\maketitle

\section{Introduction}

Currently available experimental techniques make it possible to
create two-component Bose-Einstein condensates (BEC). In the first
experiments on such systems mixtures of different hyperfine atomic
states were used, such as atoms of $^{87}$Rb with different values
of the total spin, $F$ \cite{trapp}. Such binary BEC systems offer
straightforward possibilities for the realization of novel quantum
phase transitions. In particular, while the superfluids formed by
$^{87}$Rb atoms in states $\left\vert F=1,m_{F}=-1\right\rangle $
and $\left\vert F=2,m_{F}=+1\right\rangle $ are immiscible, the
proximity of this system to the miscibility threshold makes it
possible to create specific phase-separation patterns
\cite{Rb-separation}. On the other hand, the pair of $\left\vert
F=1,m_{F}=1\right\rangle $ and $\left\vert F=2,m_{F}=2\right\rangle
$ states form a miscible superfluid, in which the separation between
the species may be induced by an external field or by a specific
flow pattern \cite{Ketterle}. The immiscibility-miscibility border
in binary systems may be shifted by means of the Feshbach-resonance
technique, which changes the strength of the nonlinear interactions
using external magnetic or optical fields \cite{Feshbach}.

Novel avenues  in the physics of multicomponent BECs were opened
with an realization of all-optical trapping and creation of spinor
condensates, where the different components correspond to the
different Zeeman levels of a given hyperfine manifold of spin $F$
(for reviews cf. \cite{vandruten,lewen-annals,Zawist}). In this case
, in principle, the number of (relevant) components can be
controlled (from $2$ to $2F+1$) using optical methods, and Feshbach
resonances may be used to tune the desired collisional channels.

 Very accurate models for the description of BEC in
quantum gases are provided by the Gross-Pitaevskii equation (GPE),
or systems of coupled GPEs, in the case of the binary system
\cite{GPE}. With the help of these equations, ground states (GSs) of
the binary systems \cite{ground} and excitations on top of them
\cite{Mason} have been studied in detail. A well-known pattern
created by the immiscibility in the binary BEC is the
domain wall between spatial regions occupied by the different species \cite%
{domain}, which may also be accurately described by means of coupled
GPEs \cite{Tripp}.

An additional physically interesting ingredient of the
binary-condensate settings is a possibility to induce the
interconversion in the mixture of two hyperfine, or Zeeman  states
of the same atom by an electromagnetic spin-flipping field
\cite{BBS}, or a two-photon Raman transition (cf. \cite{vandruten}).
In terms of the coupled GPEs, the interconversion is accounted for
by linear-coupling terms. Diverse effects have been predicted
in such models, including Josephson oscillations between the states \cite%
{Josephson}, domain walls \cite{DWs}, ``co-breathing" oscillation
modes in the mixture \cite{breathing}, nontopological vortices
\cite{vortex}, a linear-mixing-induced shift of the
miscibility-immiscibility transition \cite{Merhasin}, spontaneous
symmetry breaking in two-component gap solitons \cite{SKA}, etc.

Similar models, combining the nonlinear and linear interaction
between two modes, occur in nonlinear optics. A well-known example
is the system of coupled nonlinear Schr\"{o}dinger equations (NLSEs)
for two orthogonal polarizations of light in a nonlinear optical
fiber. In that case, the linear coupling represents the mixing
between linear polarizations in a twisted fiber, or between circular
polarizations in a fiber with an elliptically deformed core
\cite{Agrawal}. To develop a model in optics similar to that
describing the BEC, one can consider a slab waveguide with a
built-in array of fiber-like elements (strips or photonic nanowires
buried into the slab, which gives rise to the linear mixing of TE
and TM-polarized modes \cite{Panoiu}). A more versatile optical
setting may be based on slab waveguides made as a photonic crystal
\cite{phot-cryst}. Then, the propagation distance in the respective
NLSE is a counterpart of the temporal variable in the GPE, while the
transverse coordinate in the waveguide model plays the same role as
the spatial coordinate in the GPE.

An issue of obvious physical interest is the identification of the
GS in linearly-coupled binary systems. A general limitation on all
stationary states in such systems is that both components must have
equal values of the chemical potential [this condition is not
imposed onto systems with the incoherent, i.e., XPM
(cross-phase-modulation) interaction between the components].
Nevertheless, the GS in the symmetric linearly coupled system
may feature a nontrivial asymmetric shape. Typically, the \textit{%
spontaneous symmetry breaking} of the GS happens when the linear
coupling
competes with the nonlinear self-attraction acting in each component \cite%
{symmbreaking}.

The relative sign of the two stationary fields is a significant
characteristic of the GS in the linearly-coupled binary system, as
this sign couples to that of linear-mixing constant ($\Omega $)
through the corresponding term of the Hamiltonian density, hence it
affects the choice of the state providing for the minimum of the
Hamiltonian (see more details below). This circumstance is expected
to become nontrivial if $\Omega $ is subject to a spatial
modulation, which makes it a \emph{sign-changing} function of the
coordinate, $\Omega =\Omega (x)$. Indeed, if the local structure of
the GS adiabatically follows the change of $\Omega (x)$, this means
that the relative sign of the two components, being locked to the
sign
of $\Omega (x)$, must flip along with it. Therefore, each zero crossing of $%
\Omega (x)$ must give rise to a \textit{kink}, alias dark soliton,
in one of the two components, while the other one keeps a constant
sign. This idea has been developed in Ref. \cite{Dum}, where the
authors designed a method to generate dark solitons and vortices in
multicomponent BECs by means of stimulated Raman adiabatic passage
(STIRAP) from one internal state to another. While the initial BEC
had a ground state wave function (without sign changes, or
topological defects), the target BEC exhibited dark solitons or
vortices depending on the spatial dependence and symmetry of the
Raman coupling.

The objective of this work is develop the ideas of Ref. \cite{Dum}
further and identify in more detailed and systematic way the
structures of the GS in the linearly-coupled binary systems with
$\Omega (x)$ subject to the periodic change of its sign. This
situation can be readily implemented in the above-mentioned optical
model, by means of an appropriate superstructure created on top of
the slab carrying the periodic waveguiding array or the
photonic-crystal structure. In terms of the BEC, the periodic
modulation of $\Omega (x)$ may be realized too, in principle, if the
spin-flipping electromagnetic field is patterned as a standing wave.
Obviously, the latter realization cannot be achieved using a direct
(microwave) transition, as the frequency of the transition between
the hyperfine atomic states, $\sim 10$ GHz cite{BBS}, corresponds to
wavelength $\sim 30$ mm of the electromagnetic field. This is too
large in comparison with the size of experimental setups, and it
could at best allow for a {\it linear in space} coupling, that could
generate a single soliton, as pointed out in Ref. \cite{Dum}. Note,
however, that periodic modulation of $\Omega (x)$ of desired period
can be easily realized, if we use the two photon Raman transitions.

It is also worth noticing that effect discussed here have a lot of
similarity to the effect of disordered-induced-order \cite{dio},
which occurs in systems with continuous symmetry (such as complex
phase, $U(1)$ symmetry in BECs), when a symmetry breaking random,
quasi-periodic, or even periodic field (coupled to the order
parameter) is applied. Such random field will force the systems to
order in a direction "orthogonal" to the symmetry breaking field.
For example, a two-component BEC with real random (pseudo-random)
Raman coupling between the components, will order in such a way that
the relative phase between the components will be $\pi/2$
\cite{armand1}. Similar, phase control effects and disorder induced
orderings occur in Fermi superfluids \cite{armand2}, or in quantum
$XY$ spin chains in 1D \cite{armand3}.

This paper is organized as follows. The model, based on the system
of coupled NLSEs/GPEs, is formulated in Section II, where some
analytical results are reported too. In addition to the linear and
XPM couplings, the equations also include a weak trapping potential,
to make the extension of the GS effectively finite. Basic numerical
findings, in the form of actual GS profiles and maps summarizing the
most essential characteristic of the GS, \textit{viz}., the number
of kinks in it, are reported in Section III. The paper is concluded
by Section IV.

\section{The model and analytical results}

\subsection{The coupled equations}

We consider a system of equations for wave functions of the two components, $%
\psi _{1,2}$, which are cast in the notation corresponding to scaled GPEs,%
\begin{equation}
i\left( \psi _{n}\right) _{t}=\left[ -\frac{1}{2}\nabla ^{2}+V\left(
x,y\right) +\left\vert \psi _{n}\right\vert ^{2}+g\left\vert \psi
_{3-n}\right\vert ^{2}\right] \psi _{n}+\frac{1}{2}\Omega (x)\psi
_{3-n}, \label{system}
\end{equation}%
with $n=1,2$. The normalization is used to set the SPM
(self-phase-modulation) coefficients equal to $1$, while $g$ is the
free XPM coefficient. Equations (\ref{system}) are written in the
two-dimensional form, but assuming a strongly anisotropic trapping
potential, $V\left( x,y\right) =\left( 1/2\right) \left( \omega
_{x}x^{2}+\omega _{y}y^{2}\right) $, where $\omega _{x}=0.01\omega
_{y}$, and $\omega _{x}\equiv \omega =0.05$ is fixed below.
Accordingly, the equations were
solved numerically in a strongly anisotropic domain, $-25<x<+25,~-0.5<y<+0.5$%
. The modulated linear-coupling constant was taken in two forms, odd
and
even ones:%
\begin{equation}
\Omega (x)=A\left\{ \sin \left( \alpha x\right) ,\cos \left( \alpha
x\right) \right\}   \label{sincos}
\end{equation}
(most results are reported below for the former shape of the
modulation). The amplitude and wavenumber of the modulation, $A$ and
$\alpha $, will be varied below as free parameters. As concerns the
XPM coefficient, numerical results are reported below for
$g=-1,0,2/3,1,$ and $2.$ In terms of the optics, the most relevant
values are $g=2/3$ and $g=2$, which correspond,
respectively, to the pairs of linearly and circularly polarized waves \cite%
{Agrawal}. The intermediate value $g=1$ and ``exotic" ones, $%
g=0$ and $g=-1$ are, in principle, possible in photonic-crystal
slabs. In the BEC system, both positive and negative values of $g$
may be adjusted by means of the Feshbach-resonance method.

Stationary solutions to Eqs. (\ref{system}) are sought for in the
customary form, $\psi _{1,2}\left( x,y,t\right) =\exp \left( -i\mu
t\right) \phi _{1,2}(x,y)$. Real stationary wave functions $\phi
_{1,2}$ (recall chemical potential $\mu $ is identical for both
components) were found by means of
the standard numerical technique based on the solution of Eqs. (\ref{system}%
) in imaginary time \cite{imaginary}. For analytical considerations,
we adopt the stationary equations in the one-dimensional form,
\begin{equation}
\mu \phi _{n}=\left[ -\frac{1}{2}\frac{d^{2}}{dx^{2}}+\frac{\omega ^{2}}{2}%
x^{2}+\phi _{n}^{2}+g\phi _{3-n}^{2}\right] \phi
_{n}+\frac{A}{2}\left\{
\begin{array}{c}
\sin (\alpha x) \\
\cos (\alpha x)%
\end{array}%
\right\} \phi _{3-n},  \label{phi}
\end{equation}%
with $\Omega (x)$ taken as per Eq. (\ref{sincos}).

\subsection{Analytical considerations}

To understand basic properties of the GS predicted by Eqs.
(\ref{phi}), we start with the case of the uniform system
(infinitely long or subject to periodic boundary conditions), i.e.,
with $\omega =0$, and $\alpha =0$, the linear-coupling coefficient
corresponding to $\cos \left( \alpha x\right) \equiv 1$ in Eq.
(\ref{phi}). The Hamiltonian density of the uniform states
is%
\begin{equation}
\mathcal{H}=\frac{1}{2}\left( \phi _{1}^{4}+\phi _{2}^{4}\right)
+g\phi _{1}^{2}\phi _{2}^{2}+A\phi _{1}\phi _{2}.  \label{H}
\end{equation}%
A natural objective is to identify the GS as a state that minimizes
the Hamiltonian density (\ref{H}) for a fixed value of the total
atomic density,
in terms of the BEC (or total-power density, in terms of optics), $\mathcal{N%
}=\phi _{1}^{2}+\phi _{2}^{2}$.

One can find two different uniform solutions to Eqs. (\ref{phi}) with $%
\omega =\alpha =0$, symmetric and asymmetric. The former one is%
\begin{equation}
\phi _{1}=-\mathrm{sgn}(A)\phi _{2}\text{, }\phi _{1}^{2}=\phi _{2}^{2}=%
\mathcal{N}/2,~\mu =\left( 1+g\right) \mathcal{N}/2-\left\vert
A\right\vert /2,  \label{symm}
\end{equation}%
with the respective Hamiltonian density%
\begin{equation}
\mathcal{H}_{\mathrm{symm}}=\frac{1}{4}\left( g+1\right) \mathcal{N}^{2}-%
\frac{1}{2}\mathcal{N}\left\vert A\right\vert   \label{Hsymm}
\end{equation}%
[a symmetric solution subject to the opposite sign locking, $\phi _{1}=+%
\mathrm{sgn}(A)\phi _{2}$, exists too, but it gives rise to a higher
Hamiltonian density]. The asymmetric solution is%
\begin{equation}
\phi _{1,2}^{2}=\frac{1}{2}\left[ \mathcal{N}\pm \sqrt{\mathcal{N}%
^{2}-\left( g-1\right) ^{-2}A^{2}}\right] ,~\mu =\mathcal{N}
\label{asymm}
\end{equation}%
($+$ and $-$ pertain to $\phi _{1}$ and $\phi _{2}$), with the
relative sign of the wave functions locked so as to satisfy
$\mathrm{sgn}(\phi _{1}\phi _{2})=-\mathrm{sgn}\left( \left(
g-1\right) A\right) $. The corresponding
values of the Hamiltonian density is%
\begin{equation}
\mathcal{H}_{\mathrm{asymm}}=\frac{1}{2}\mathcal{N}^{2}-\frac{1}{4}\frac{%
A^{2}}{g-1}.  \label{Hasymm}
\end{equation}

Obviously, asymmetric solution (\ref{asymm}) exists under the condition of $%
\left\vert g-1\right\vert >|A|/\mathcal{N}$, and the comparison of
expressions (\ref{Hsymm}) and (\ref{Hasymm}) demonstrates that the
GS, which must realize the minimum of the Hamiltonian density,
corresponds to the asymmetric state at $g>1$, and to the symmetric
one at $g<1$. This result exactly coincides with the elementary
immiscibility (miscibility) condition in the uniform medium, $g>1$
($g<1$) \cite{Mineev}. Thus, in the ideal uniform configuration, the
point of the transition between the symmetric and asymmetric states,
$g=1$, is not shifted by the linear mixing. However, the difference
from the linearly uncoupled system is that, in the immiscible phase
(at $g>1$), the binary system without the linear coupling must form
domain walls , while the linearly coupled system may stay uniform,
spontaneously concentrating a larger number of atoms (or larger
optical power, in terms of optics) in one component.

The state with a strongly broken symmetry, i.e., $\left\vert \phi
_{2}\right\vert \ll \left\vert \phi _{1}\right\vert $, which
corresponds to small values of $|A|$, can be found as an approximate
\emph{nonuniform} analytical solution to the full system of
equations (\ref{phi}), which include the trapping potential (
$\omega >0$), and $\alpha \neq 0$. In the zero-order approximation
($A=0$), one may take the solution as $\phi _{2}=0$ and $\phi
_{1}(x)$ in the form of the Thomas-Fermi (TF)\ ansatz. After that,
the first-order ($\sim A$) solution for $\phi _{2}(x)$, and the
solution for $\phi _{1}(x)$, which includes the second-order ($\sim
A^{2}$) correction, are found in the following form: at
$x^{2}<2\tilde{\mu}/\omega ^{2}$,
\begin{equation}
\phi _{2}(x)=-\frac{A}{2}\frac{\sqrt{\tilde{\mu}-\omega
^{2}x^{2}/2}}{\left(
g-1\right) \left( \tilde{\mu}-\omega ^{2}x^{2}/2\right) +\alpha ^{2}/2}%
\left\{
\begin{array}{c}
\sin (\alpha x) \\
\cos (\alpha x)%
\end{array}%
\right\} ,  \label{phi2}
\end{equation}%
\begin{equation}
\phi _{1}(x)=\sqrt{\tilde{\mu}-\omega ^{2}x^{2}/2}\left\{ 1\mp \frac{A^{2}}{%
16}\frac{\cos \left( 2\alpha x\right) }{\left( \alpha ^{2}+\tilde{\mu}%
-\omega ^{2}x^{2}/2\right) \left[ \left( g-1\right) \left( \tilde{\mu}%
-\omega ^{2}x^{2}/2\right) +\alpha ^{2}/2\right] }\right\} ,
\label{phi1}
\end{equation}%
and $\phi _{1}=\phi _{2}=0$ at $x^{2}>2\tilde{\mu}/\omega ^{2}$. Here, $%
\sqrt{\tilde{\mu}-\omega ^{2}x^{2}/2}$ is the usual form of the TF
approximation for $\phi _{1}(x)$, with the effective chemical
potential including a small correction, $\tilde{\mu}\equiv \mu
+A^{2}/\left( 4\alpha ^{2}\right) $, which eliminates a formal
secular term at the second order of the perturbative expansion. The
upper and lower signs in front of $A^{2}$ in Eq. (\ref{phi1})
correspond, respectively, to the upper and lower rows in Eqs.
(\ref{phi}) and (\ref{phi2}).

\section{Numerical results}

\subsection{Ground-state profiles}

Simulations of Eqs. (\ref{system}) in imaginary time $t\equiv -i\tau
$ were performed in the interval of $0<\tau <500$, with time step
$\Delta \tau
=10^{-3}$. In fact, the convergence to stationary patterns was achieved by $%
\tau \sim 100-200$. The finally generated patterns do not depend on
the choice of the seed configuration in the imaginary-time
integration.

\begin{figure}[tph]
\centering\includegraphics[width=18.0cm]{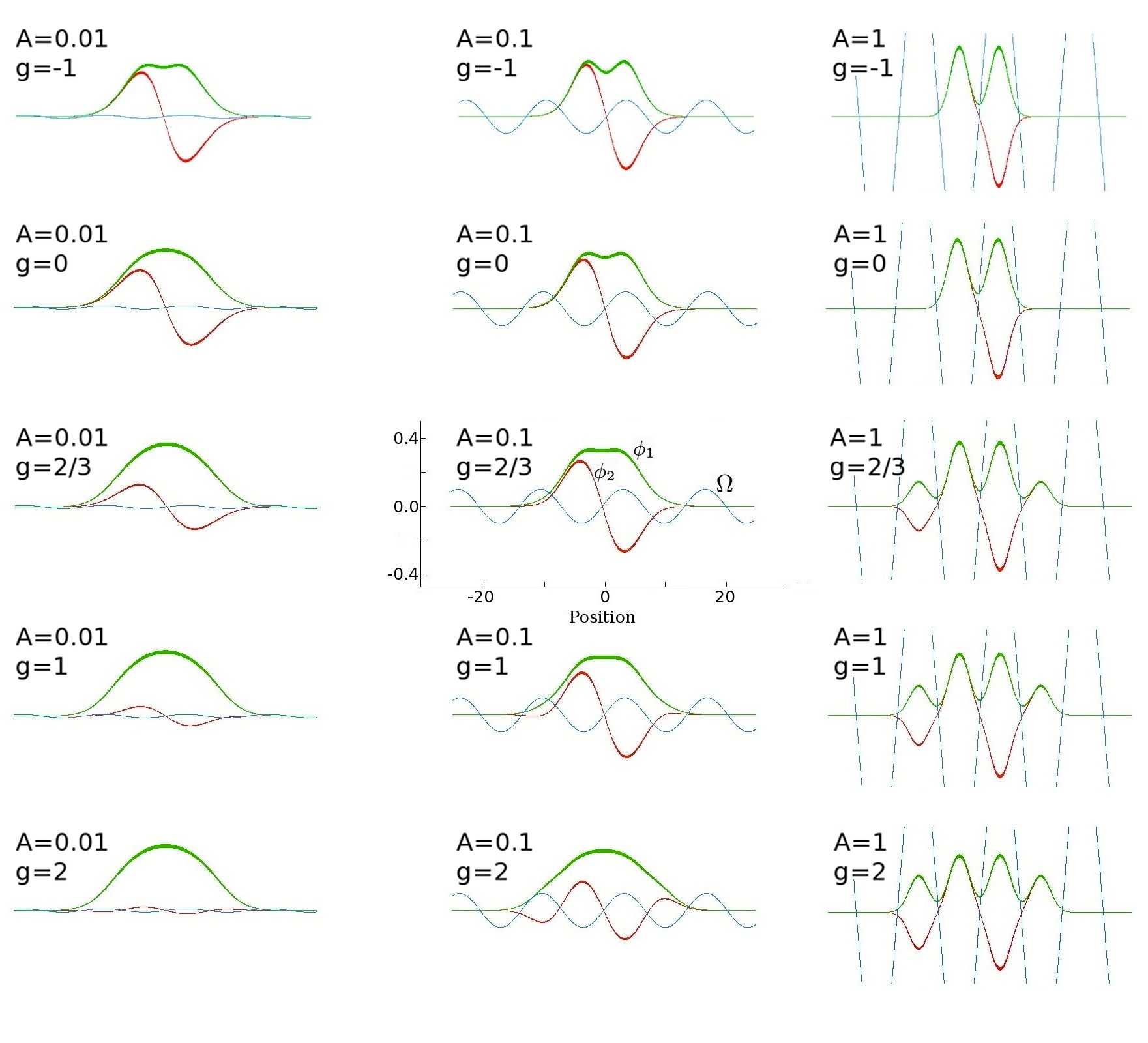}
\caption{(Color online) Stationary wave functions $\protect\phi
_{1}(x)$ (green) and $\protect\phi _{2}(x)$ (red) in the case of the
odd modulation function, $\Omega (x)=A\sin \left( \protect\alpha
x\right) $ (blue), see Eq. (\protect\ref{sincos}). The left, middle
and right columns correspond to the
weak, moderate, and strong linear coupling, respectively, with $A=0.01$, $%
A=0.1$, and $A=1$. The rows running from top to bottom correspond
the following values of the XPM coefficient: $g=-1,0,2/3,1,2$.}
\label{fig1}
\end{figure}

A set of typical profiles of the GSs generated by the numerical
method are displayed in Fig. \ref{fig1}. The profiles were obtained
varying the amplitude of the odd modulation function in Eq.
(\ref{phi}), $A$, and the XPM coefficient, $g$, for a particular
fixed value of the modulation wavenumber, $\alpha =0.4624$ (other
values $\alpha \lesssim 1$ produce essentially the same picture, see
also below).

The approximate analytical solution given by Eqs. (\ref{phi2}),
(\ref{phi1}) adequately describes the general shape of the patterns,
including the facts that the number of kinks is odd in the case of
the odd modulation [and even, in the case of the even modulation, as
shown by the numerical results generated for function $\cos \left(
\alpha x\right) $ in Eq. (\ref{phi})]. The dip in the profile of the
stationary wave function $\phi _{1}(x)$, which
is observed, around $x=0$, at $g=-1$ for $A=0.01$, and at $g=-1,0,2/3$ for $%
A=0.1$, is correctly explained by Eq. (\ref{phi1}), which includes
the correction to the TF (zero-order) approximation with the
curvature at the center opposite to that of the TF waveform, the
amplitude of the correction growing with the decrease of $g$.
Further, the numerically found wave
function $\phi _{2}(x)$ is accurately approximated by Eq. (\ref{phi2}) at $%
g=2/3,1,2$ for $A=0.01$, and at $g=2$ for $A=0.1$, the agreement
being qualitative in other cases.

Comparing these results to analytical solutions (\ref{symm}) and (\ref{asymm}%
) obtained in the uniform system, we conclude that, at $g=-1$ and $g=0$ for $%
A=0.01,$ at $g=-1,0,2/3,1$ for $A=0.1,$ and all values of $g$ for
$A=1$, both components in the numerically found solutions have
approximately equal amplitudes, the difference from the uniform
system being that the periodic change of the sign of $\Omega (x)$
gives rise to the array of kinks in $\phi _{2}(x)$. Another
essential deviation from the uniform system is observed in the
transition between the states with nearly equal and strongly
different amplitudes of the two components: as shown above, in the
uniform system the transition point, $g=1$, does not depend on the
linear-coupling coefficient, $A$, while the numerical results
demonstrate a strong dependence of the transition on $A$: in the
case of the weak coupling, $A=0.01$, the transition occurs around
$g=0$, where the GS of the uniform system would remain symmetric; in
the case of the moderate coupling, $A=0.1$, the transition takes
place around $g=2/3$, i.e., close to the point $g=1$ predicted by
the analysis of the uniform system; finally, the strong linear
coupling, $A=1$, keeps the amplitudes of the components virtually
equal even at $g=2,$ where the uniform system would feature the
strong asymmetry, as per solution (\ref{asymm}).

\subsection{The number of kinks in the ground states}

The number of kinks is the most essential overall (topological)
characteristic of the GS [we stress that the kinks are always formed
in one component, which is denoted $\phi _{2}(x)$, while the other
component, $\phi _{1}(x)$, does not change its sign]. As said above,
in the case of the
odd/even spatial modulation of the linear coupling in Eqs. (\ref{phi}) and (%
\ref{sincos}), the number of kinks is always odd/even too. Analyzing
numerical data, we counted those kinks which showed, in their core
areas, the amplitude of $\phi _{2}$ that would be no smaller than
$5\%$ of the largest value of $\psi _{1}$. In terms of the present
analysis, this actually means that the kinks were taken into account
if the respective amplitude of $\phi _{2}(x)$ would not fall below
$\phi _{2}=0.02$.

\begin{figure}[tph]
\centering\includegraphics[width=14.0cm]{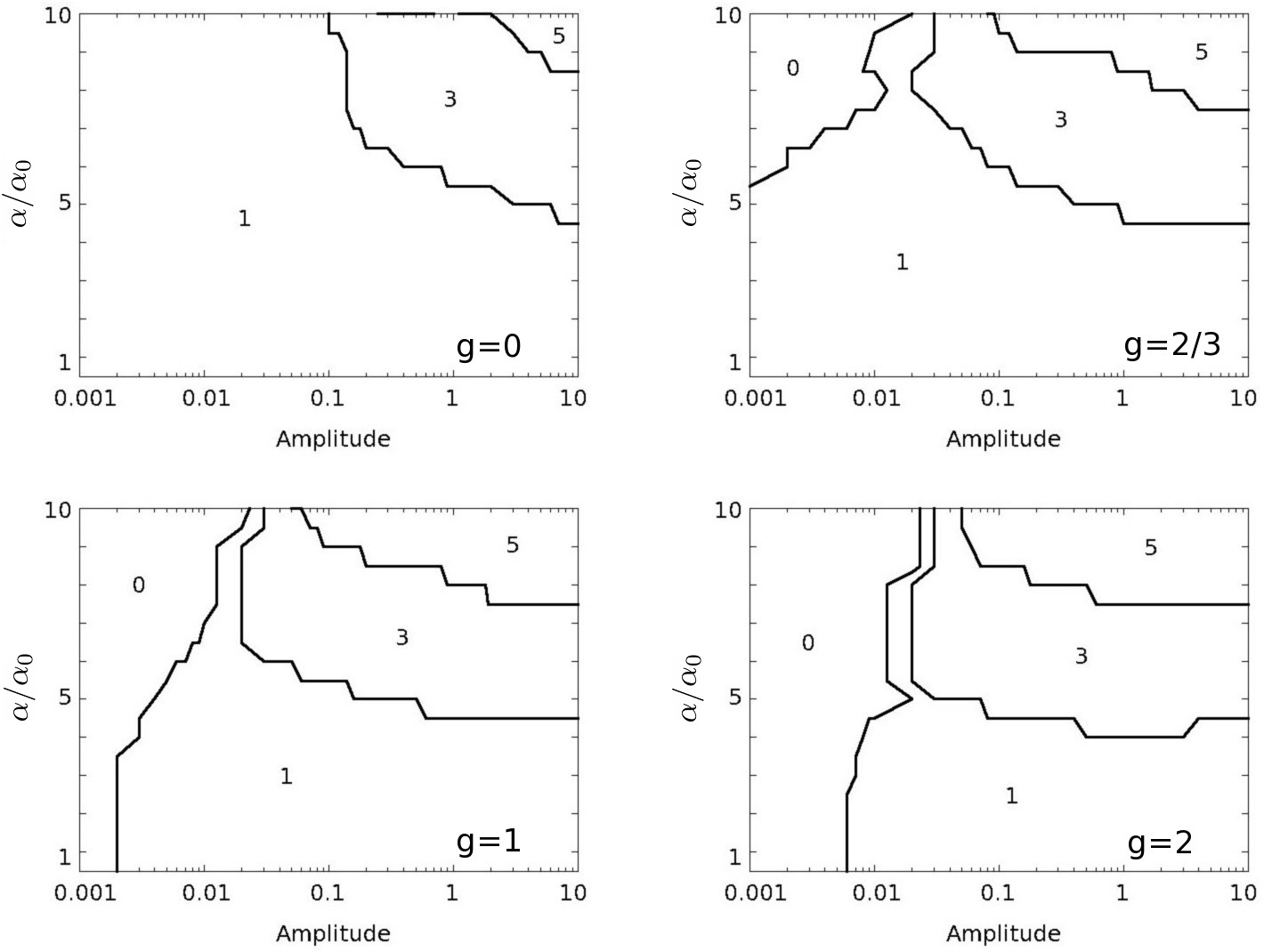}
\caption{
The number of kinks in the numerically found
ground states, as a function of amplitude $A$ and normalized
wavenumber $\protect\alpha/\alpha_0$ (with $\alpha_0=0.09248$) of
the odd spatial modulation of the linear coupling, see Eq.
(\protect\ref{sincos}). The maps are displayed at four fixed values
of the XPM coefficients, $g=0,2/3,1,2$. The values of $A$ are shown
on the logarithmic scale, to comprise the cases of weak, moderate,
and strong linear coupling. In this case of the odd modulation, the
number of kinks is odd too, see the text.} \label{fig2}
\end{figure}

The number of kinks in the GS is presented, as a function of the
linear-coupling strength $A$ and modulation wavenumber $\alpha $, in
maps
collected in Fig. \ref{fig2}, at four fixed values of the XPM coefficient $g$%
. The case of $g=-1$ is not included, as it always gives rise to a
single kink, cf. Fig. \ref{fig1}. The latter observation is
explained by the fact
that the \emph{attraction} between the two components, in the case of $g=-1$%
, impedes $\phi _{2}(x)$ to cross zero while $\phi _{1}(x)$ is not
too small, except for at the central point. On the other hand, at
$g>0$ the repulsion between the components facilitates the
appearance of the zero crossings of $\phi _{2}(x)$, thus leading to
the increase in the number of kinks. We note that the maps
pertaining to different positive values of $g$, i.e., $g=2/3,$ $1$,
and $2$, are not drastically different.

Quite naturally, the number of the kinks increases with the
modulation wavenumber, $\alpha $, as the kinks are associated with
points where $\Omega (x)$ changes its sign. The same argument
explains why the number of kinks increases too with the modulation
strength, $A$.

\begin{figure}[tph]
\centering\includegraphics[width=8.0cm]{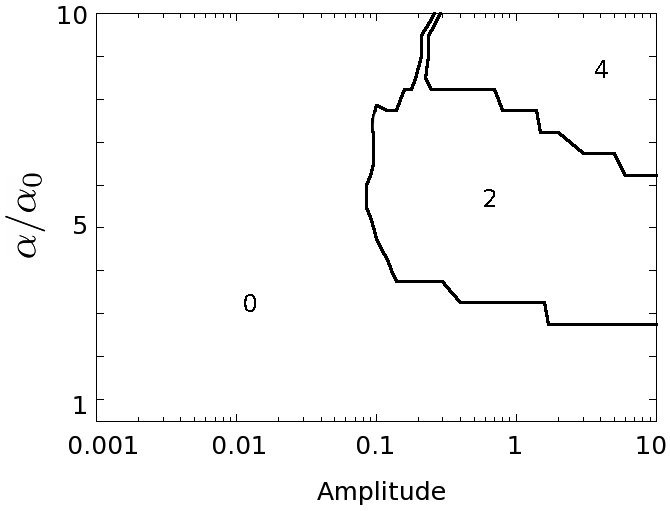} \caption{
The same as in Fig. \protect\ref{fig3} (again, with
$\alpha_0=0.09248$), but for the case of the even modulation
function, $\cos \left( \protect\alpha x\right) $, in Eq.
(\protect\ref{sincos}), and $g=0$. The number of kinks is even in
the case of the even modulation, see the text.} \label{fig3}
\end{figure}

Some data concerning the number of kinks has also been collected for
the case of the even modulation of the local modulation, with
function $\cos \left( \alpha x\right) $ in Eq. (\ref{sincos}). The
respective map is displayed in Fig.~\ref{fig3} for $g=0$. The
absence of the kink at the central point makes their overall number
smaller than in the case of the odd modulation. In this case,
general features exhibited by the dependence of the number of kinks
on parameters $A,\alpha $ and $g$ are qualitatively the same as in
the case of the odd modulation.

\section{Conclusion}

The objective of this work was to identify the GSs (ground states)
in two-component models based on the system of two NLSEs/GPEs
coupled by the linear interconversion and nonlinear (XPM)
interaction, with the self-repulsion acting in each component. A
well-known property of such systems is that their GSs may be
represented by either symmetric or asymmetric states, depending on
the strengths of linear and XPM couplings. In the present model, we
have extended the idea of Ref. \cite{Dum} and considered in detail
the periodic odd, or even spatial modulation of the linear coupling,
$\Omega (x)$. The repeated change of the sign by $\Omega (x)$
produces a strong impact on the structure of the GS in the
two-component system, as the relative sign of the components tends
to be locked to the local sign of $\Omega $. Accordingly, the GS of
the trapped system features one or several kinks in one component,
while the other component does not change its sign. The parity of
the number of kinks coincides with that of the modulation function.
The interplay of the modulation of $\Omega (x)$ with the trapping
potential also strongly affects the transition between the states
with nearly equal and strongly different amplitudes of the two
components. These results were obtained by means of the systematic
numerical analysis, and also in the analytic perturbative form. The
model may be realized in nonlinear optics, in terms of periodically
modulated slab waveguides. Also, it may also be implemented in BECs,
when two-photon Raman transitions are used.

A challenging problem is to extend the analysis to two-dimensional
binary systems, with a two-dimensional (checkerboard-patterned)
modulation of the linear coupling. This two-dimensional setting may
also be realized in the photonic-crystal medium. In  2D, as as
already discussed in \cite{Dum}, it should be possible to generate
vortices and (in more complex settings) other types of topological
defects. In fact, this method of vortex generation is completely
analogue to the methods of creating "artificial" magnetic fields
employing spatially dependent Berry's phase \cite{Juze},  or Raman
couplings \cite{Lin}. Another interesting extension of the present
studies is to perform a similar analysis for quantum systems,
formulated in terms of an appropriate Bose-Hubbard model, i.e. in
the strongly correlated limit, where one expects the appearance of
quantum solitons (cf. \cite{kostia} and references therein).

B.A.M. appreciates hospitality of ICFO (Institut de Ci\`{e}ncies Fot\`{o}%
niques), Castelldefels (Barcelona), Spain. The work of this author
was supported, in a part, by the German-Israel Foundation through
grant No. 149/2006. We acknowledge also financial support from the
Spanish MINCIN project FIS2008-00784 (TOQATA), Consolider Ingenio
2010 QOIT, EU STREP project NAMEQUAM, ERC Advanced Grant QUAGATUA,
 and from the Humboldt Foundation.

\end{document}